\documentclass[a4paper,11pt]{article}
\usepackage{pos}

\newcommand\be{\begin{equation}}
\newcommand\ee{\end{equation}}
\newcommand{\bea}{\begin{eqnarray}}
\newcommand{\eea}{\end{eqnarray}}
\newcommand{\nn}{\nonumber}
\newcommand{\pd}{\partial}

\title{Hidden Symmetries, Rapid Turns and Cosmic Acceleration}
\author*{Lilia Anguelova}

\affiliation{Institute for Nuclear Research and Nuclear Energy,\\
  Bulgarian Academy of Sciences,\\ Tsarigradsko Chaussee 72, Sofia 1784, Bulgaria}

\emailAdd{anguelova@inrne.bas.bg}

\abstract{Hidden symmetries provide a powerful tool for finding exact solutions in multifield cosmological models. We review how, using such symmetries, one can find inflationary solutions in two-field models, which lead to the generation of primordial black holes. We also discuss an exact solution in a two-field cosmological model, which describes dark energy. This solution is obtained with the use of a hidden symmetry, although the latter is broken by a constant term in the scalar potential. All of the above solutions are characterized by field-space trajectories with rapid turns.}

\FullConference{%
  
11th International Conference of the Balkan Physical Union (BPU11),\\
  28 August - 1 September 2022\\
  Belgrade, Serbia
}

\begin{document}
\maketitle

\section{Introduction}

According to current observational data, ordinary matter and radiation make up only a small fraction of the energy density of the Universe today. The dominant components are the so called dark energy and dark matter, whose nature is not well understood. This motivates the investigation of a variety of cosmological models beyond the standard $\Lambda$CDM one. Among them, models with multiple scalar fields are particularly natural from the perspective of compatibility with quantum gravity \cite{GK,OPSV,AP,BPR}. Such multifield models are of interest both for understanding cosmological inflation in the early Universe, as well as for describing dark energy in the late Universe.

The equations of motion of these models, however, are rather involved generically. So it is common practice to investigate them by using numerical methods. Needless to say, knowing the background solutions analytically could be very beneficial for gaining better understanding of the relevant physics. The search for analytical solutions is greatly facilitated by the presence of hidden symmetries. This is because the latter can lead to significant simplification of the field equations, thus enabling one to find exact solutions. However, hidden symmetries can exist only under certain conditions, which impose restrictions both on the form of the scalar potential and on the geometry of the scalar field space. 

The conditions for hidden symmetry, in a certain type of two-field cosmological models, were studied in detail in \cite{ABL,ABL2}. Considering a particular family of symmetries, \cite{ABL} found many classes of exact solutions of the corresponding background equations of motion. One of those classes leads to inflationary models with large enough perturbations to induce the formation of primordial black holes \cite{LA}, which are of interest as a component of dark matter. Hidden symmetry is also crucial for obtaining an exact solution that gives a two-field dark energy model \cite{ADGW}, even though the relevant potential breaks the symmetry by a constant term.

We review the main results of \cite{ABL,ABL2}, focusing on a particular hidden symmetry relevant for the background solutions studied in \cite{LA} and \cite{ADGW}. Then we describe the properties of the class of exact solutions that leads to primordial black hole generation. In particular, we explain the behavior of the corresponding field-space trajectories, which are characterized by a single rapid turn. Further, we discuss the derivation of the class of exact solutions that behave as dark energy. The field-space trajectories of those solutions are inward spirals (toward the center of field space), characterized by an always-large turning rate. Despite that, the consistency conditions of \cite{AL}, for long-term rapid turn inflation, do not apply to these dark energy solutions, as we explain in concluding remarks.

\section{Two-field cosmological models with hidden symmetry}
\setcounter{equation}{0}

We will consider cosmological models, which arise from the minimal coupling of a set of scalar fields $\phi^I (x^{\mu})$ to Einstein gravity. This system is described by the action:
\be \label{Action_gen}
S = \int d^4x \sqrt{-\det g} \left[ \frac{R}{2} - \frac{1}{2} G_{IJ} \pd_{\mu} \phi^I \pd^{\mu} \phi^J - V (\{ \phi^I \}) \right] \,\,\, ,
\ee
where $g_{\mu \nu}$ is the spacetime metric with $\mu,\nu = 0,1,2,3$ and $G_{IJ}$ is the metric on the scalar field-space with coordinates $\{\phi^I\}$\,. Our focus will be the two-field case, in which ${I,J = 1,2}$\,. The standard Ansatze for a spatially homogeneous cosmological background are:
\be \label{metric_g}
ds^2_g = -dt^2 + a^2(t) d\vec{x}^2 \qquad , \qquad \phi^I = \phi^I_0 (t) \quad ,
\ee 
where $ds^2_g$ is the spacetime metric with scale factor $a(t)$\,. Also, as usual, the Hubble parameter is:
\be
H (a) = \frac{\dot{a}(t)}{a(t)} \,\,\, .
\ee

The equations of motion for the background fields $\phi^I_0$ and $a$\,, which follow from the above action, are a rather complicated coupled system that often can be solved only numerically. However, certain classes of exact solutions can be found by using hidden symmetries. The idea is the following. Substituting the background Ansatze (\ref{metric_g}) in the action (\ref{Action_gen}) gives (after integration by parts) the Lagrangian density:
\be \label{Ac_aphi}
{\cal L} = - 3 a \dot{a}^2 + a^3 \!\left[ \frac{1}{2} G_{IJ} \dot{\phi}_0^I \dot{\phi}_0^J - V (\phi_0) \right] \,\,\, ,
\ee
per unit spatial volume. Viewing ${\cal L}$ as a classical mechanical Lagrangian for the generalized coordinates $q^{\cal{I}} \equiv \{ a, \phi_0^I \}$\,, we can look for Noether symmetries of the system. For that purpose, let us consider coordinate transformations $q^{\cal I} \rightarrow Q^{\cal I} (q)$ generated by a vector field of the form:
\be \label{XaXI}
X = X^a (a,\phi_0) \pd_a + X^I (a,\phi_0) \pd_{\phi_0^I} \,\,\, .
\ee
They induce corresponding transformations, generated by:
\be
\hat{X} = X + \dot{X}^a \pd_{\dot{a}} + \dot{X}^I \pd_{\dot{\phi}_0^I} \,\,\, ,
\ee
on the manifold with coordinates $\{ q^{\cal I}, \dot{q}^{\cal I} \}$\,. The Lagrangian ${\cal L} (q^{\cal I}, \dot{q}^{\cal I})$ has a Noether symmetry when:
\be \label{Lie_der}
L_{\hat{X}} {\cal L} = 0 \,\,\, ,
\ee
where $L_{\hat{X}}$ is the Lie derivative along $\hat{X}$\,. Finding solutions of (\ref{Lie_der}) enables one to simplify the Euler-Lagrange equations of (\ref{Ac_aphi}) by transforming to suitable generalized coordinates, which are adapted to the symmetry. This facilitates greatly the search for exact background solutions.\footnote{The Euler-Lagrange equations of (\ref{Ac_aphi}) are equivalent with the original equations of motion arising from (\ref{Action_gen}), when the Hamiltonian constraint $E_{\cal L} \equiv (\pd_{\dot{q}^{\cal I}} \!{\cal L}) \,\dot{q}^{\cal I} - {\cal L} = 0$ is imposed. The effect of this constraint is to reduce by one the number of independent integration constants of the solutions; see for instance \cite{CdeR}. \label{HamC}} Note that the set of conditions, contained in (\ref{Lie_der}), not only determines the symmetry generators, but also constrains the form of the scalar potential $V(\{\phi_0^I\})$\,.

The above symmetry conditions were studied in detail in \cite{ABL2}. It was shown there that (\ref{Lie_der}) implies the following general form for the components $X^a$ and $X^I$ of the symmetry generator (\ref{XaXI}):
\bea
X^a (a, \phi_0) &=& \frac{\Lambda (\phi_0)}{\sqrt{a}} \quad , \nn \\ 
X^I (a, \phi_0) &=& Y^I (\phi_0) \,\, - \,\, 4 \,\frac{G^{IJ} (\phi_0) \,\pd_J \Lambda (\phi_0)}{a^{3/2}} \quad ,
\eea
where $\pd_I \!\equiv \!\pd_{\phi_0^I}$ and $Y^I (\phi_0)$ is any field-space Killing vector, which preserves the potential $V (\phi_0)$\,, while $\Lambda (\phi_0)$ is an arbitrary function satisfying the equation:
\be \label{Eq_G-Lambda}
\left( \pd_I \pd_J - \Gamma^K_{IJ} \pd_K \right) \Lambda \,= \,\, \frac{3}{8} \,G_{IJ} \Lambda \quad .
\ee
Here $\Gamma^K_{IJ}$ are the Christoffel symbols of the field-space metric $G_{IJ}$\,. In addition, it also follows from (\ref{Lie_der}) that the scalar potential has to satisfy:
\be \label{VLambda_eq}
G^{IJ} \pd_I V \pd_J \Lambda \,= \,\frac{3}{4} \,V \Lambda \quad ,
\ee 
in order to be compatible with the existence of such a symmetry \cite{ABL2}. Note that symmetries with nonvanishing $\Lambda$, called {\it hidden} symmetries, mix the scale factor $a$ and the scalar fields $\phi_0^I$\,. Hence, they are present only at the level of the reduced Lagrangian (\ref{Ac_aphi}), not the full action (\ref{Action_gen}). Thus, while such hidden symmetries are very useful for finding background solutions, they do not affect the action for the perturbations around those backgrounds.

For the purposes of the subsequent sections, we will be interested in a special case of the above considerations, which is given by choosing $G_{IJ}$ to be the metric on the Poincar\'e disk. In that case, and using for convenience the notation:
\be \label{Backgr_id}
\phi^1_0 (t) \equiv \varphi (t) \qquad {\rm and} \qquad \phi^2_0 (t) \equiv \theta (t) \,\,\, ,
\ee
we can write the field-space metric as (see, for instance, \cite{ABL}): 
\be \label{Gmetric}
ds^2_{G} = d\varphi^2 + f(\varphi) d\theta^2 \,\,\, ,
\ee
where\footnote{Note that the general form of the function $f$ for the Poincar\'e disk case is $f(\varphi) = \frac{1}{q^2} \sinh^2 (q \varphi) $ with arbitrary $q$\,. However, in (\ref{f_Pdisk}) we have fixed $q = \sqrt{\frac{3}{8}}$\,, as required for the existence of a hidden symmetry \cite{ABL,ABL2}.}
\be \label{f_Pdisk}
f (\varphi) = \frac{8}{3} \,\sinh^2 \left( \sqrt{\frac{3}{8}} \,\varphi \right) \,\,\, .
\ee
With $G_{IJ}$ given by (\ref{Gmetric})-(\ref{f_Pdisk}), the Lagrangian (\ref{Ac_aphi}) has the form:  
\be \label{L_class_mech}
{\cal L} \,= \,- 3 a \dot{a}^2 + \frac{a^3 \dot{\varphi}^2}{2} + \frac{a^3 f (\varphi) \,\dot{\theta}^2}{2} - a^3 V(\varphi,\theta) \,\,\, .
\ee
The most general solution of the $\Lambda$-equation (\ref{Eq_G-Lambda}), resulting from the metric (\ref{Gmetric})-(\ref{f_Pdisk})\,, is \cite{ABL2}:
\be \label{Lambda-Pd_sol}
\Lambda (\varphi , \theta) = C_0 \,\cosh \left( \sqrt{\frac{3}{8}} \,\varphi \right) + (C_1 \sin \theta + C_2 \cos \theta) \,\sinh \left( \sqrt{\frac{3}{8}} \,\varphi \right) \,\,\, ,
\ee
where $C_{0,1,2} = const$.\footnote{The $\Lambda$-solution in (\ref{Lambda-Pd_sol}) with $C_0 = 0$ was obtained earlier in \cite{ABL}.} Using (\ref{Lambda-Pd_sol}) in (\ref{VLambda_eq}), one can determine the most general form of the scalar potential $V (\varphi , \theta)$\,, compatible with this hidden symmetry \cite{ABL2}. 

The particular symmetry, which will be of interest in the next sections, is obtained for:
\be \label{C0C1C2}
C_0 = 0 \quad , \quad C_1 = 1 \quad , \quad C_2 = 0 \quad .
\ee
In that case, the potential $V(\varphi, \theta)$ simplifies considerably (see also \cite{ABL}). Specializing further to rotationally invariant potentials, the form of $V$, compatible with the hidden symmetry given by (\ref{Lambda-Pd_sol})-(\ref{C0C1C2}), is the following \cite{ABL}: 
\be \label{PotExSym}
V (\varphi) = V_0 \,\cosh^2 \left( \sqrt{\frac{3}{8}} \,\varphi \right) \,\,\, ,
\ee
where $V_0 = const > 0$. For $\Lambda (\varphi , \theta)$ of the form (\ref{Lambda-Pd_sol})-(\ref{C0C1C2}), the generalized coordinate transformation from $(a, \varphi , \theta)$ to symmetry-adapted coordinates $(u,v,w)$ can be written as \cite{ABL}:\footnote{Here and in the following, we use the more convenient notation of \cite{LA}. In addition, we have introduced explicitly the constant $\theta_0$ for future use.}
\bea \label{Ch_var}
a (t) &=& \left[ u^2 (t) - \left( v^2 (t) + w^2 (t) \right) \right]^{1/3} \,\,\,\, , \nn \\
\varphi (t) &=& \sqrt{\frac{8}{3}} \,{\rm arccoth} \!\left( \sqrt{\frac{u^2 (t)}{v^2 (t) + w^2 (t)}} \,\,\right) \,\,\,\, , \nn \\
\theta (t) &=& \theta_0 + {\rm arccot} \!\left( \frac{v (t)}{w (t)} \right) \,\,\,\, , \,\,\,\, \theta_0 = const \quad .
\eea
Substituting (\ref{Ch_var}) in (\ref{L_class_mech}) simplifies greatly the kinetic terms:
\be \label{LkinV}
{\cal L} \,= \,- \frac{4}{3} \dot{u}^2 + \frac{4}{3} \dot{v}^2 + \frac{4}{3} \dot{w}^2 - a^3 V \,\,\, .
\ee
For the potential (\ref{PotExSym}), we also have: 
\be \label{a3V_sym}
a^3 V = V_0 u^2 \,\,\, .
\ee
%However, we have not specified the form of $V$ in (\ref{LkinV}) in anticipation of the considerations of Section \ref{DE}. 
Clearly, the Euler-Lagrange equations of (\ref{LkinV}) are much simpler to solve than the original field equations.

\section{Transient rapid turning and primordial black holes} \label{PBH}
\setcounter{equation}{0}

In this Section we will focus on a particular class of exact solutions of the equations of motion, resulting from (\ref{Action_gen})-(\ref{metric_g}). These background solutions were found in \cite{ABL} by taking $G_{IJ}$ and $V$ as in (\ref{Gmetric})-(\ref{f_Pdisk}) and (\ref{PotExSym}), respectively, and using the hidden symmetry generated by (\ref{Lambda-Pd_sol})-(\ref{C0C1C2}). The form of the exact solutions is given by (\ref{Ch_var}), together with:
\bea \label{Sols_uvw}
u (t) &=& C^u_1 \sinh \!\left( \kappa \,t \right) \,+ \,C^u_0 \cosh \!\left( \kappa \,t \right) \quad , \quad \kappa \equiv \frac{1}{2} \sqrt{3 V_0} \quad \,, \nn \\
v (t) &=& C_1^v \,t \,+ \,C_0^v \quad \,, \nn \\
w (t) &=& C_1^w t \,+ \,C_0^w \quad ,
\eea
where $C^{u,v,w}_{0,1} = const$\,, as well as the constraint: 
%To ensure that the Einstein equations (\ref{EinstEq}) are also satisfied, one has to impose the following constraint among the integration constants in (\ref{Sols_uvw}):
\be \label{Constr_s}
(C_1^v)^2 + (C_1^w)^2 \,= \,\kappa^2 \left[ (C_1^u)^2 - (C_0^u)^2 \right] \,\,\, .
\ee
Relation (\ref{Constr_s}) arises from imposing the constraint, discussed in footnote \ref{HamC}, on the Lagrangian (\ref{LkinV}). Note that, by redefining the integration constants, one can absorb $\theta_0$ of (\ref{Ch_var}) into the constants $C_0^{v,w}$ of (\ref{Sols_uvw}), thus obtaining exactly the form of the expressions in \cite{ABL,LA}.

The background solutions, given by (\ref{Ch_var}) and (\ref{Sols_uvw})-(\ref{Constr_s}), exhibit a key property, which is necessary to induce the generation of primordial black holes. To explain that, let us define several important characteristics of an inflationary background trajectory in field space. To begin with, it is convenient to introduce tangent and normal vectors to a trajectory $(\phi^1_0(t),\phi^2_0(t))$\, as follows:
\be \label{TNbasis}
T^I = \frac{\dot{\phi}^I_0}{\dot{\sigma}} \quad , \quad N_I = (\det G)^{1/2} \epsilon_{IJ} T^J \quad , \quad \dot{\sigma}^2 \equiv G_{IJ} \dot{\phi}^I_0 \dot{\phi}^J_0 \,\,\,\, .
\ee
These vectors form an orthonormal basis. Using it, one can define the turning rate of a trajectory as \cite{AAGP}:
\be \label{Om}
\Omega = - N_I D_t T^I \quad , \quad D_t T^I \equiv \dot{\phi}^J_0 \nabla_J T^I \,\,\, .
\ee
A nonvanishing turning rate function \,$\Omega (t)$ signifies a deviation of the background  trajectory from a geodesic. Three other important functions, characterizing an inflationary solution, are given by:
\be \label{SR_par}
\varepsilon = - \frac{\dot{H}}{H^2} \qquad , \qquad \eta_{\parallel} = - \frac{\ddot{\sigma}}{H \dot{\sigma}} \qquad , \qquad \eta_{\perp} = \frac{\Omega}{H} \quad .
\ee
Clearly, $\varepsilon$ and $\eta_{\parallel}$ are the usual slow roll parameters for a single-field model with inflaton $\sigma (t)$\,. The quantity $\eta_{\perp}$\,, on the other hand, is a new distinguishing feature of two-field models. For more details on (\ref{SR_par}), see for instance \cite{LA}, or the concise summary in \cite{LA2}.

\begin{figure}[t]
%\begin{figure}[h!]
\begin{center}
\hspace*{-0.2cm}
\includegraphics[scale=0.33]{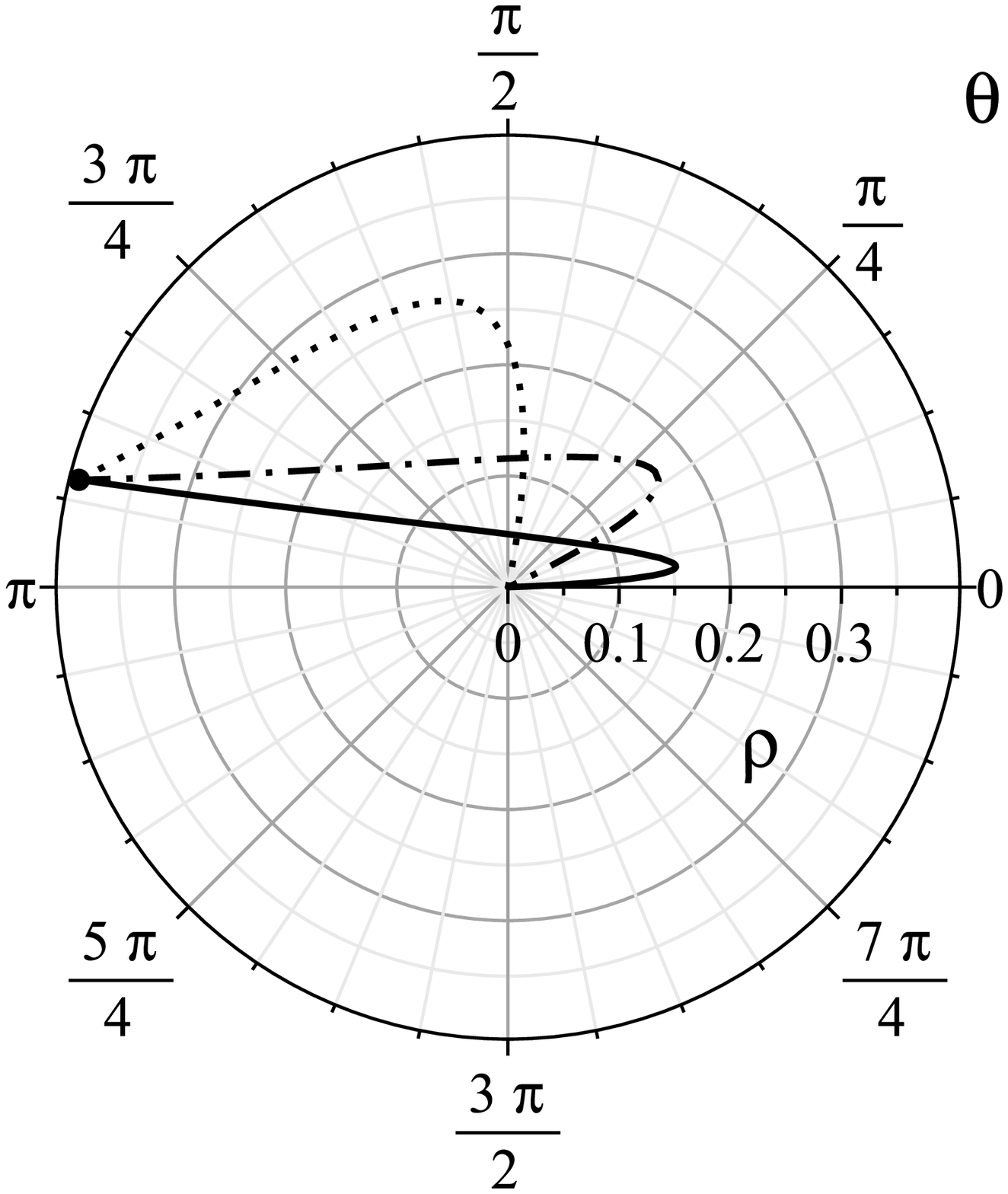}
\hspace*{0.3cm}
\includegraphics[scale=0.33]{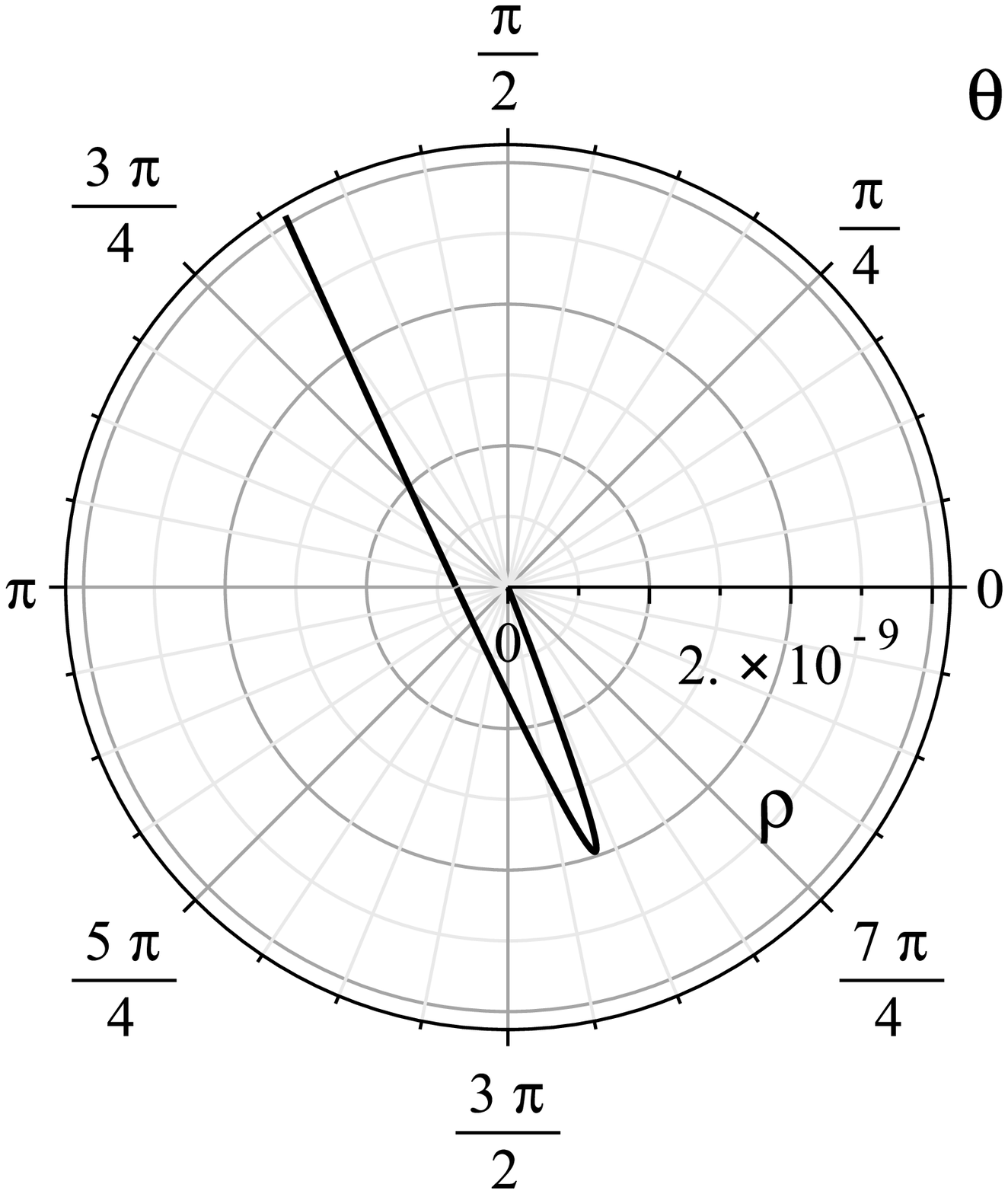}
\end{center}
\vspace{-0.6cm}
\caption{{\small Examples of trajectories $\left( \rho (t), \theta (t) \right)$ of the exact solutions given by (\ref{Ch_var}) and (\ref{Sols_uvw})-(\ref{Constr_s}), where $\rho \equiv \tanh \!\left( \frac{\sqrt{6}}{8} \varphi \right)\in [0,1)$ and $\theta_0 \equiv 0$\,. \underline{{\it On the left}}: We have taken $C^u_1=3$ , $C^u_0=\frac{3}{2}$ , $C^v_0=-1$ , $C^w_0=\frac{1}{4}$ together with $C^w_1=\frac{1}{8}$ ({\it solid line}) , $C_1^w = 3$ ({\it dash-dotted line}) and $C_1^w = 7.5$ ({\it dotted line})\,; in all three cases $C_1^v$ is determined by the positive root of (\ref{Constr_s}). The starting point of the trajectories at $t=0$ is the solid dot at $(\rho,\theta)\approx\left( \,0.4\,,\frac{15}{16} \,\pi \,\right)$\,. \underline{{\it On the right}}: We have taken $C^u_0=6$ , $C^v_1=-\frac{1}{5}$ , $C^v_0=1$ , $C^w_1=\frac{1}{2}$ , $C_0^w = -2.47$ and $C_1^u$ is determined by the positive root of (\ref{Constr_s}). This plot starts at a later time than $t=0$\,, since otherwise the shape of the trajectory would not be visible as \,$\rho|_{t=0} = 0.23 \,>\!\!> \,\rho|_{\Omega_{peak}} \approx 2 \times 10^{-9}$\,.}}
\label{ShapeTraj}
\vspace{0.1cm}
\end{figure}

Computing the turning rate of a background trajectory, for $G_{IJ}$ as in (\ref{Gmetric}) and $\pd_{\theta} V = 0$ as in (\ref{PotExSym}), one obtains \cite{LA}:
\be \label{Om_PD}
\Omega = \frac{\sqrt{f}}{\left( \dot{\varphi}^2 + f \dot{\theta}^2 \right)} \,\dot{\theta} \,\pd_{\varphi} V \,\,\, .
\ee
Then, substituting (\ref{f_Pdisk}) and (\ref{PotExSym})-(\ref{Ch_var}) in (\ref{Om_PD}) gives:
\be \label{Om_uvw}
\Omega \,= \frac{3V_0}{4} \,\frac{u \,(v \dot{w} - \dot{v} w) \,\sqrt{u^2-w^2-v^2}}{\left[ (v \dot{u} - \dot{v} u)^2 + (w \dot{u} - \dot{w} u)^2 - (v \dot{w}-\dot{v} w)^2 \right]} \,\,\, .
\ee
Writing down explicitly the further substitution of (\ref{Sols_uvw}) in (\ref{Om_uvw}) is not illuminating, as that leads to a rather messy expression. Nevertheless, \cite{LA} managed to show that the resulting function $\Omega (t)$ has a single peak, whose height can be varied at will by choosing appropriately the values of the integration constants. Before and after the peak, the turning rate tends to zero. This is precisely the shape of $\Omega (t)$ necessary for primordial black hole (PBH) generation according to \cite{PSZ,FRPRW}. The peak of the turning rate (\ref{Om_uvw}) is due to a single turn of the corresponding field space trajectory. We have illustrated the shape of the background trajectories, obtained from (\ref{Ch_var}) and (\ref{Sols_uvw})-(\ref{Constr_s}), in Figure \ref{ShapeTraj} by using for convenience the canonical radial coordinate on the Poincar\'e disk $\rho \in [0,1)$\,, which is related to $\varphi$ in (\ref{Gmetric})-(\ref{f_Pdisk}) via $\rho = \tanh \!\left( \frac{\sqrt{6}}{8} \varphi \right)$ \cite{ABL}. On the left of the Figure, we have plotted three examples, which show how the shape of a trajectory can be varied by varying the integration constants; sharper turns correspond to higher peaks of \,$\Omega (t)$ \cite{LA}. These examples are convenient for illustrating the entire shape of the trajectory. However, in them $\varepsilon (t)$ is not necessarily small et early times (although it tends to zero with $t$) and, in addition, the rapid turn occurs within the first e-fold of inflationary expansion. On the other hand, the example on the right of Figure \ref{ShapeTraj} illustrates a typical trajectory with slow roll $\varepsilon$-parameter satisfying \,$\varepsilon (t) <\!\!< 1$ always\footnote{Indeed, it was shown in \cite{LA} that, for the exact solutions under consideration, the phenomenologically desirable slow roll regime is obtained in a neighborhood of the origin of the Poincar\'e disk, i.e. for $\rho <\!\!< 1$ and thus $\varphi <\!\!< 1$\,. Note that this is unlike the $\alpha$-attractor models of \cite{KLR,KLR2,KL3,CKLR}, for which slow roll occurs near the boundary, i.e. for $\varphi \rightarrow \infty$\,.}, as well as with a turning-rate peak occurring after tens of e-folds of expansion. The entire duration of the rapid turn, however, is contained within a single e-fold, for the whole class of exact solutions. 

Physically, the sharp rapid turn of a background trajectory can lead to significant enhancement of the power spectrum of the curvature perturbation $\zeta$ (and, hence, of the density fluctuations), because the strength of interaction between $\zeta$ and the entropic perturbation around the trajectory depends on \,$\eta_{\perp} \!= \Omega / H$ \,(for more details on this mechanism, see \cite{PSZ,FRPRW}). The entropic perturbation arises from the following decomposition of the scalars around the background:
\be \label{Decomp}
\phi^I (t, \vec{x}) = \phi^I_0 (t) + \delta \phi^I (t, \vec{x}) \,\,\, .
\ee
In terms of the basis (\ref{TNbasis}), one can expand the fluctuations in (\ref{Decomp}) as:
\be
(\delta \phi)^I = (\delta \phi)_{\parallel} T^I + (\delta \phi)_{\perp} N^I \,\,\, .
\ee
The adiabatic perturbation \,$(\delta \phi)_{\parallel}$ \,can be gauged away (in comoving gauge: $\!(\delta \phi)_{\parallel} \equiv 0$), leaving only the entropic scalar perturbation $(\delta \phi)_{\perp}$ as physical degree of freedom. The turning rate \,$\Omega$ \,enters in the expression for the effective mass of $(\delta \phi)_{\perp}$ as follows (see, for instance, \cite{AAGP}):
\be \label{Meff_entropic}
M_{(\delta \phi)_{\perp}}^2 = N^I N^J V_{;IJ} - \Omega^2 + \varepsilon H^2 {\cal R} \,\,\, ,
\ee
where \,$V_{;IJ} = \pd_I \pd_J V - \Gamma^K_{IJ} \,\pd_K V$ \,and \,${\cal R}$ \,is the Ricci scalar of \,$G_{IJ}$\,. This implies that high enough peaks of \,$\Omega$ \,(as needed for PBH generation) can induce tachyonic instability of the entropic perturbation. It was shown in \cite{LA} that, indeed, the rapid turn of a background trajectory, determined by (\ref{Ch_var}) and (\ref{Sols_uvw})-(\ref{Constr_s}), leads to a brief period with \,$M_{(\delta \phi)_{\perp}}^2 \!(t) < 0$ \,and that, accordingly, a higher peak of the turning rate gives rise to a greater magnitude of the tachyonic instability. The value of \,$-M_{(\delta \phi)_{\perp}}^2 \!(t)$ \,at the peak of \,$\Omega (t)$ \,can be made as large as necessary by making suitable choices for the values of the integration constants.

We have seen above that the parameters $\varepsilon (t)$ and $\eta_{\perp} (t)$ of the exact solutions, given by (\ref{Ch_var}) and (\ref{Sols_uvw})-(\ref{Constr_s}), have the desired behavior to induce PBH generation. However, the remaining inflationary parameter $\eta_{\parallel}$ in (\ref{SR_par}), computed for those solutions, behaves in a phenomenologically problematic manner.\footnote{Recall that $\eta_{\parallel}$ coincides with the Hubble slow roll parameter $\eta_H \!\equiv \!- \frac{\ddot{H}}{2 H \dot{H}}$ on solutions of the background equations of motion.} So we would like to modify the above exact solutions in a way that improves the behavior of $\eta_{\parallel} (t)$\,, without spoiling the rest of their properties. Such a modification does, indeed, exist as was shown in \cite{LA}. The class of modified (approximate) solutions, found there analytically, arises from a one-parameter deformation of (\ref{f_Pdisk}) and (\ref{PotExSym}). For generic values of that new parameter, the hidden symmetry (\ref{Lambda-Pd_sol})-(\ref{C0C1C2}) is broken, although it is recovered for a particular value. Under certain condition on the new parameter, the modified non-symmetric solutions preserve the desired for PBH-generation behavior of $\eta_{\perp}$\,, described above, while they have $ \eta_{\parallel} \approx 3$ before the peak of the turning rate and $\eta_{\parallel} \approx 0$ after it. Thus, the rapid-turn period in these solutions represents a transition between an ultra-slow roll and a slow roll inflationary phases. For more details see \cite{LA}.

\section{Long-lasting rapid turning and dark energy} \label{DE}
\setcounter{equation}{0}

In this Section we will be interested in background trajectories, whose turning rate is large for a prolonged (or unlimited) period of time, and corresponding models of multifield dark energy. To explain how hidden symmetry can, again, be of crucial use for finding analytically such solutions, let us start by writing down explicitly the relevant background equations of motion.

As in the previous sections, we will use the identifications (\ref{Backgr_id}) and will consider a rotationally invariant field space metric, in which case $G_{IJ}$ can be written as (\ref{Gmetric}) with arbitrary positive-definite $f (\varphi)$\,. Then, (\ref{Action_gen})-(\ref{metric_g}) lead to the following field equations:
\bea
\ddot{\varphi} - \frac{f'}{2} \dot{\theta}^2 + 3 H \dot{\varphi} + \pd_{\varphi} V = 0 \quad &,& \quad \ddot{\theta} + \frac{f'}{f} \dot{\varphi} \dot{\theta} + 3 H \dot{\theta} + \frac{1}{f} \pd_{\theta} V = 0  \quad , \label{ScalarEoMs}\\
\dot{\varphi}^2 + f \dot{\theta}^2 = - 2 \dot{H} \quad &,& \quad 3H^2 + \dot{H} = V  \quad . \label{EinstEqf}
\eea
Here (\ref{ScalarEoMs}) are the equations of motion for the background scalars, while (\ref{EinstEqf}) are the Einstein equations. To preserve the $U(1)$ isometry of (\ref{Gmetric}), we will again consider only $\theta$-independent potentials. Thus, in the following, we will take: 
\be \label{V-th_ind}
\pd_{\theta} V \equiv 0 \,\,\, .
\ee
Now, since we are interested in solutions of (\ref{ScalarEoMs})-(\ref{EinstEqf}), whose field-space trajectories exhibit long-lasting rapid tuning, we can make the Ansatz:
\be \label{dotTh_const}
\dot{\theta} \equiv \omega = const \,\,\, .
\ee
Physically, if the condition $\dot{\theta} \approx const$ is not satisfied, then the turning rate is unlikely to remain large for very long. So generic solutions, which have a prolonged rapid turning period, are likely to be well approximated during that period by exact solutions with (\ref{dotTh_const}). The ansatz (\ref{dotTh_const}) underlies the proposal of \cite{ASSV} for  multifield dark energy.

In \cite{ADGW} equations (\ref{ScalarEoMs})-(\ref{EinstEqf}) were analyzed carefully, under the assumptions (\ref{V-th_ind})-(\ref{dotTh_const}), and it was shown that solutions exist only if certain relation between the functions $f$ and $V$ is satisfied. For a given $f(\varphi)$\,, this relation is an ODE, which determines the form of $V (\varphi)$ compatible with such solutions. In general, this ODE is rather involved. However, it simplifies considerably for some specific choices of field-space metric. In particular, taking $f$ as in (\ref{f_Pdisk}), one obtains a simple equation, whose solution is \cite{ADGW}: 
\be \label{Vs}
V (\varphi) \,\, = \,\, C_V \,\cosh^2 \left( \frac{\sqrt{6}}{4} \,\varphi \right) \,- \,\frac{4}{3} \,\omega^2 \quad , \quad C_V = const \,\,\, .
\ee
Note that, to guarantee that this potential is always positive, one has to impose:
\be \label{CvO}
C_V > \frac{4}{3} \,\omega^2 \ \,.
\ee
The functional form of (\ref{Vs}) is exactly the same as (\ref{PotExSym}), which was determined by the hidden symmetry (\ref{Lambda-Pd_sol})-(\ref{C0C1C2}). However, the constant $\omega$-term in (\ref{Vs}) breaks that symmetry, since it is incompatible with (\ref{VLambda_eq}). Despite that, the symmetry-adapted transformation (\ref{Ch_var}) again leads to significant simplification. Indeed, substituting (\ref{f_Pdisk}), (\ref{Vs}) and (\ref{Ch_var}) in (\ref{L_class_mech}) gives:
\be \label{Lagr_uvw}
{\cal L} \,= \,- \frac{4}{3} \dot{u}^2 + \frac{4}{3} \dot{v}^2 + \frac{4}{3} \dot{w}^2 - \frac{4}{3} \kappa^2 u^2 - \frac{4}{3} \omega^2 v^2 - \frac{4}{3} \omega^2 w^2 \,\,\,\, ,
\ee
where
\be \label{kappa_def}
\kappa \equiv \frac{1}{2} \sqrt{3C_V - 4 \omega^2} \,\,\, .
\ee
Of course, (\ref{Lagr_uvw}) has the same kinetic terms as (\ref{LkinV}). Now, though, the potential term contains all variables, unlike (\ref{a3V_sym}), since there is no symmetry and, thus, no cyclic variable anymore. 

Solving the Euler-Lagrange equations of (\ref{Lagr_uvw}) is rather easy. Using the resulting solutions inside the $\theta (t)$-expression in (\ref{Ch_var}) and then imposing (\ref{dotTh_const}) fixes some of the integration constants. So one obtains:
\be \label{uvw_Sols_DE}
u (t) = C_1^u \sinh (\kappa t) + C_0^u \cosh (\kappa t) \quad , \quad v (t) = C_w \cos (\omega t) \quad , \quad w (t) = C_w \sin (\omega t) \quad ,
\ee
where $C^u_{0,1}\,, C_w = const$\,. Substituting (\ref{uvw_Sols_DE}) in (\ref{Ch_var}) gives:
\bea \label{Sol_om}
a (t) &=& \left( u^2 (t) - C_w^2 \right)^{1/3} \,\,\,\, ,  \quad u (t) = C_1^u \sinh (\kappa t) + C_0^u \cosh (\kappa t) \,\,\,\, , \nn \\
\varphi (t) &=& \sqrt{\frac{8}{3}} \,{\rm arccoth} \!\left( \sqrt{\frac{u^2 (t)}{C_w^2}} \,\,\right) \,\,\,\, , \nn \\
\theta (t) &=& \theta_0 + \omega t \,\,\,\, .
\eea
The analogue of the constraint (\ref{Constr_s}) is now \cite{ADGW}:
\be \label{Constr_sDE}
\kappa^2 \!\left[ (C_0^u)^2 - (C_1^u)^2 \right] + 2 \omega^2 C_w^2 = 0 \,\,\,\, .
\ee
Expressions (\ref{Sol_om})-(\ref{Constr_sDE}), together with (\ref{kappa_def}), solve the background equations of motion (\ref{ScalarEoMs})-(\ref{EinstEqf}), when $f$ and $V$ are given by (\ref{f_Pdisk}) and (\ref{Vs}) respectively.

This class of exact solutions was investigated in detail in \cite{ADGW}, where it was shown that it provides a model of dark energy. In particular, the corresponding equation-of-state parameter tends to $-1$ with time, for any physically allowed values of the integration constants.\footnote{Requiring that $a(t)$ and $\dot{a}(t)$ are real and positive restricts the choices of values for the integration constants \cite{ADGW}.} Also, the Hubble parameter of these solutions approaches fast a constant, whereas their $\varepsilon (t)$ parameter is a monotonically decreasing function that tends to zero. Hence, the spacetime of the solutions (\ref{Sol_om})-(\ref{Constr_sDE}) approaches de Sitter space more and more closely with time. So, as classical backgrounds, they represent cosmological models that are rather similar to $\Lambda$CDM. However, the behavior of scalar perturbations around the background can lead to significant differences, as discussed in \cite{ASSV}.

%\begin{figure}[h!]
\begin{figure}[t]
\begin{center}
\hspace*{-0.2cm}
\includegraphics[scale=0.33]{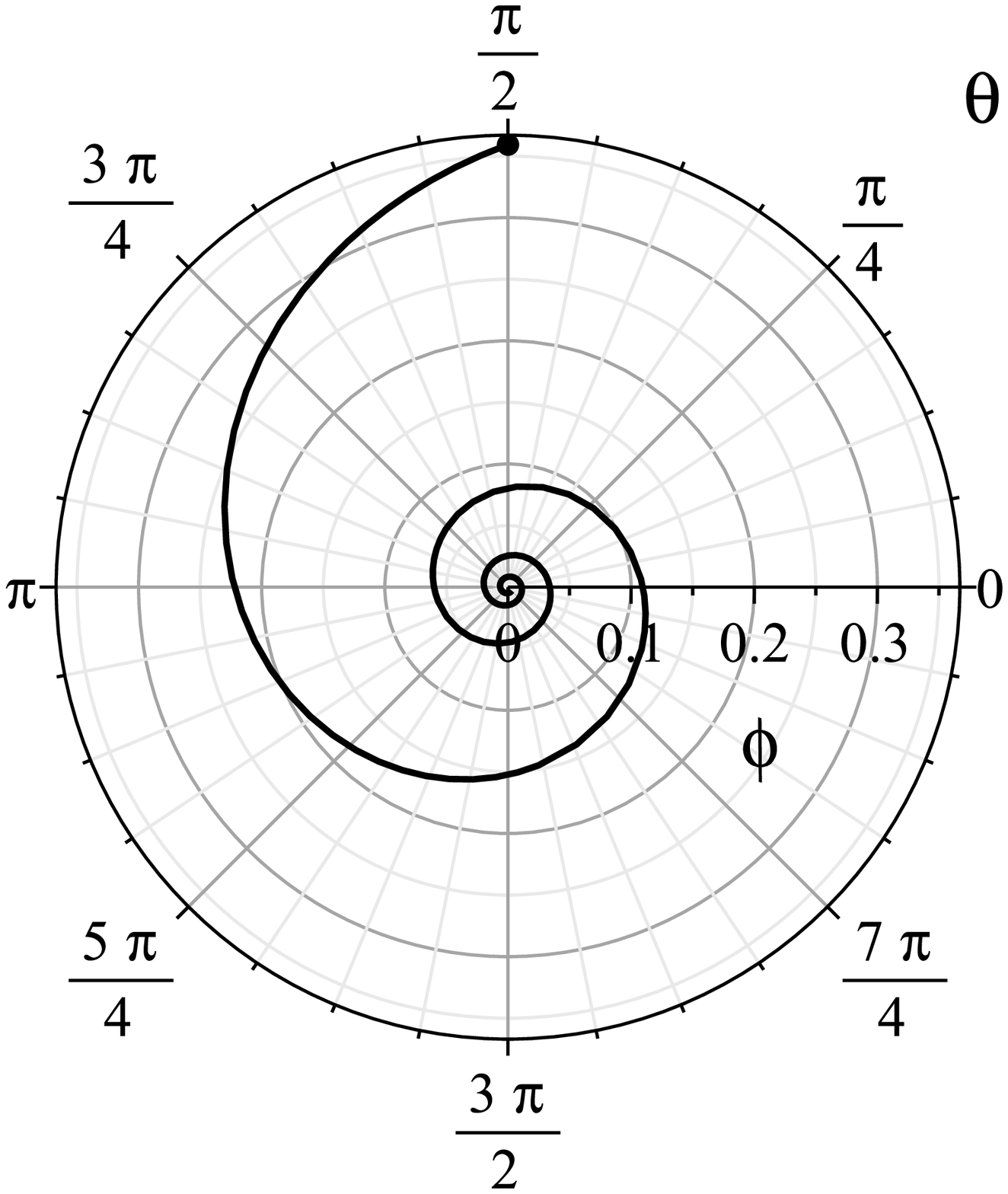}
\hspace*{0.2cm}
\includegraphics[scale=0.33]{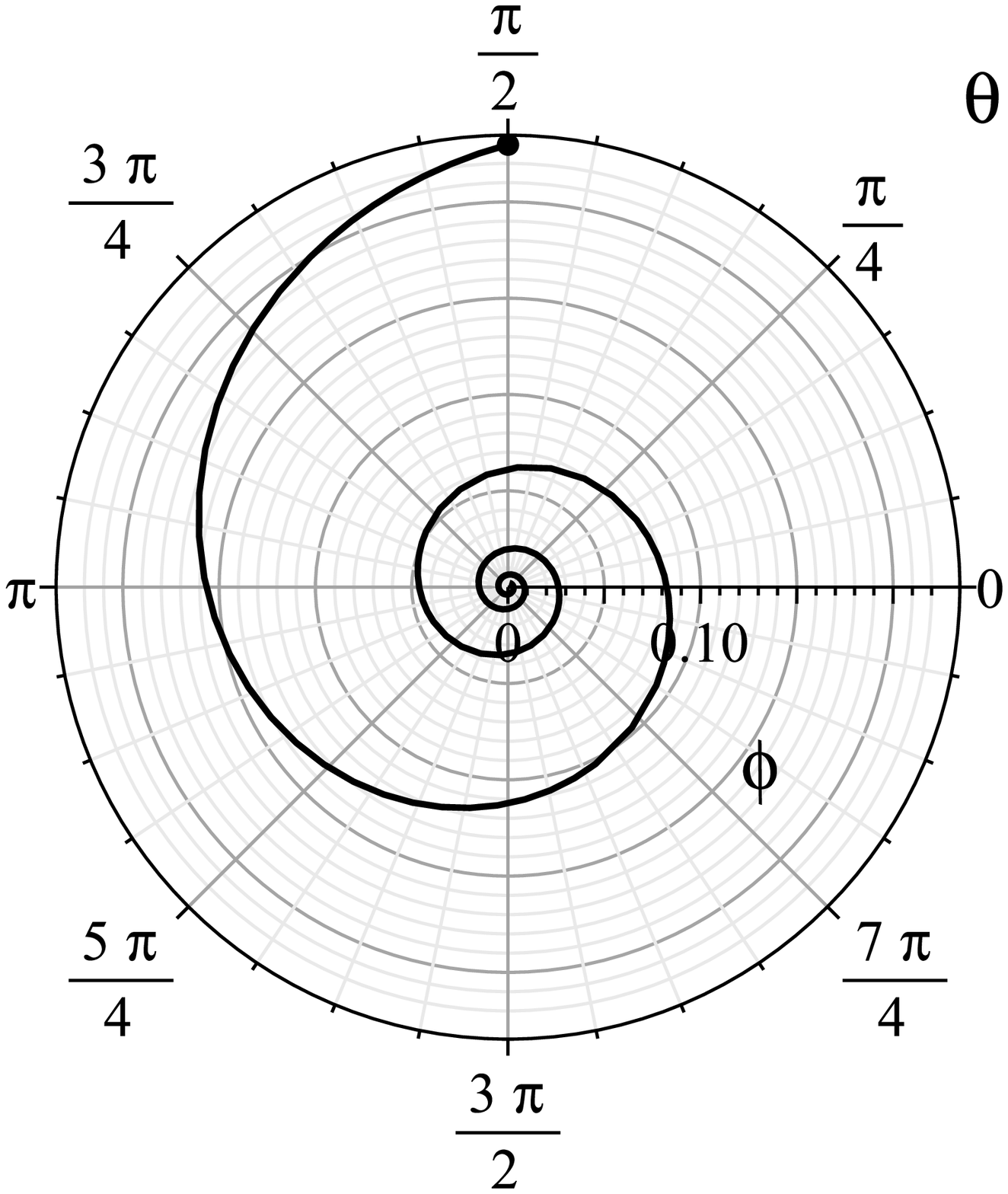}
\end{center}
\vspace{-0.6cm}
\caption{{\small Two examples of trajectories $\left(\varphi(t), \theta(t)\right)$ of the exact solutions (\ref{Sol_om})-(\ref{Constr_sDE}), together with (\ref{kappa_def}). We have taken $ \omega= 4$ and $C_V = 22$\,, as well as $C^u_0 = 1$\,, $C^u_1 = 2$ ({\it on the left}) and $C^u_0 = 2$\,, $C^u_1 = 3$ ({\it on the right}). In both cases $C_w$ is determined by the positive root of (\ref{Constr_sDE}). The starting point of the trajectories at $t=0$ is denoted by a solid dot, situated at $\theta = \frac{\pi}{2}$ and $\varphi \neq 0$\,.}}
\label{TrajDE}
\vspace{0.1cm}
\end{figure}

The field-space trajectories of the exact solutions (\ref{Sol_om})-(\ref{Constr_sDE}), together with (\ref{kappa_def}), are spirals toward the center (at $\varphi = 0$) of the Poincar\'e disk with metric (\ref{Gmetric})-(\ref{f_Pdisk}). Equivalently, they are spirals toward the minimum of the potential (\ref{Vs})-(\ref{CvO})\,. We have illustrated the shape of these trajectories with two examples in Figure \ref{TrajDE}. The numerical values of the constants, in these examples, are such that the rapid-turning condition $\eta_{\perp}^2 (t) \!>\!\!> \!1$ is always satisfied, where $\eta_{\perp}$ was introduced in (\ref{SR_par}). Note that, despite having an always-large turning rate, these solutions do not develop a tachyonic instability of the kind discussed in Section \ref{PBH}. More precisely, one can show that now the effective mass (\ref{Meff_entropic}) of the entropic perturbation always satisfies \cite{ADGW}:
\be \label{Meff_pos}
M_{(\delta \phi)_{\perp}}^2 \!(t) \,> \,0 \,\,\, ,
\ee
as long as $\varepsilon <\!\!< 1$\,.\footnote{In that regard, note that the scalar curvature of the metric (\ref{Gmetric})-(\ref{f_Pdisk}) is constant, namely ${\cal R}=-3/4$\,. So the $\varepsilon$-term in (\ref{Meff_entropic}) is negligible during slow roll.} In fact, (\ref{Meff_pos}) is satisfied also for values of the integration constants in (\ref{Sol_om})-(\ref{Constr_sDE}), such that initially one has $\varepsilon|_{t=0} \sim {\cal O}(1)$\,. But even in those cases, just as for any other physically allowed values of the constants, the value of the monotonically decreasing parameter $\varepsilon (t)$ fast becomes exceedingly small with increasing $t$.

The slow roll parameter $\eta_{\parallel}$\,, however, is not small for this class of background solutions. In fact, $\eta_{\parallel} (t)$ is always of ${\cal O} (1)$ and tends fast to $\frac{3}{2}$ with time. Since this dark energy parameter is not constrained by current observations, the violation of the slow roll condition $\eta_{\parallel} \!<\!\!< \!1$ is not an issue here, unlike for models of inflation. However, it has the important consequence that the consistency conditions of \cite{AL}, which are necessary to ensure sustained rapid turning in the slow roll regime, do not apply to the solutions of this Section.

\section*{Acknowledgements}

I am grateful to E. M. Babalic, J. Dumancic, R. Gass, C. Lazaroiu, P. Suranyi and L.C.R. Wijewardhana for valuable discussions and collaboration. I have received partial support from the Bulgarian NSF grant KP-06-N38/11.

\end{document}